\documentclass[pre,floats,superscriptaddress,showpacs,usenames]{revtex4}
\usepackage{amsmath}
\usepackage{graphicx}

\usepackage[dvipsnames]{color}

\usepackage{epsfig}

\newcommand{\bea}{\begin{eqnarray}}
\newcommand{\eea}{\end{eqnarray}}
\newcommand{\be}{\begin{equation}}
\newcommand{\ee}{\end{equation}}

\newcommand{\bm}{\mathbf}

\newcommand{\bk}{\hat{\bf k}}
\newcommand{\sab}{\sigma^{\alpha\beta}}
\newcommand{\gab}{g^{\alpha\beta}}
\newcommand{\muab}{\mu^{\alpha\beta}}

\newcommand{\bfxi}{\mbox{\boldmath $\xi$}}

\newcommand{\xx}{{\bf x}}
\newcommand{\rr}{{\bf r}}
\newcommand{\NN}{{\bf \nabla}}
\newcommand{\FF}{{\bf F}}

\newcommand{\vv}{{\bf v}}
\newcommand{\cc}{{\bf c}}
\newcommand{\uu}{{\bf u}}
\newcommand{\uua}{{\bf u}^\alpha}

\renewcommand{\ss}{{\bf\sigma}}

\newcommand{\shat} {\hat{\ss}}

\newcommand{\bfnabla}{\mbox{\boldmath $\nabla$}}
\newcommand{\bnabri}{{\bfnabla_{{\bf r}_i}}}

\newcommand{\na}{n^\alpha}
\newcommand{\nb}{n^\beta}

\newcommand{\ma}{m^\alpha}
\newcommand{\mb}{m^\beta}

\newcommand{\fvec}{{\bf F}}

\begin{document}
\date{\today}

\title{Kinetic Density Functional Theory:\\ A microscopic approach to fluid mechanics}

\author{Umberto Marini Bettolo Marconi\footnote[3]
{(umberto.marinibettolo@unicam.it)}
}

\address{ Scuola di Scienze e Tecnologie, 
Universit\`a di Camerino, Via Madonna delle Carceri, 62032 ,
Camerino, INFN Perugia, Italy}

\author{Simone Melchionna}
\address{Istituto Processi Chimico-Fisici, Consiglio Nazionale delle Ricerche, Italy}
\begin{abstract}
In the present paper we give  a brief summary of some recent theoretical advances in the treatment
of inhomogeneous fluids and methods which have applications in the study of dynamical properties
of liquids in situations of extreme confinement, such as nanopores, nanodevices, etc.
The approach obtained by  combining  kinetic and density functional methods is microscopic, fully self-consistent and
 allows  to determine both configurational 
 and  flow properties of dense fluids. 
 The theory predicts the correct hydrodynamic behavior and
provides a practical and numerical tool to determine how the transport properties are modified
when the length scales of the confining channels  are comparable with the size of the molecules.
The  applications range from the dynamics of simple fluids under confinement,
to that of neutral binary mixtures and electrolytes where the theory  in the limit of slow gradients
reproduces the known phenomenological equations such as the Planck-Nernst-Poisson
and the Smolochowski equations.
The approach here illustrated allows for fast
numerical solution of the evolution equations for the one-particle phase-space distributions  by means of 
the weighted density lattice Boltzmann method and is particularly useful when one considers  flows in complex geometries.
  \end{abstract}

\maketitle

\section{Introduction}

One of
the goals of statistical physics is to bridge different levels of descriptions
of matter accounting for macroscopic, mesoscopic and microscopic levels.
A complete description at the
finest scale makes in principle possible to derive the properties at the grosser scales using
a reduction process usually carried out only through a series of
 approximations \cite{castiglione2008chaos,pavliotis2008multiscale,weinan2011principles}.

In Thermodynamics and  Hydrodynamics, which both refer to the macroscopic level,  
a different strategy is adopted and instead of using a reduction process, one deduces 
relations between a reduced number of macroscopic variables such as temperature, density, energy, fluid velocity, etc.
from experimental observations and/or phenomenological arguments  \cite{huang2001introduction,batchelor2000introduction}
.
  
 In principle one may obtain information about large assemblies of interacting molecules via molecular dynamics (MD) 
numerical simulations, but the method requires considerable computational time and computer memory
especially in systems where one is interested to explore properties relative to species having very low concentration.
Alternatively,
the theoretical methods which can be used to study the dynamical behavior of fluid solutions  
are 
the  DDFT (dynamic density functional theory), 
which describes an  overdamped dynamics typical of colloidal behavior 
\cite{marconi1999dynamic,archer2004dynamical,rauscher2010ddft}  the
 WDLBM (weighted density Lattice Boltzmann method)
 which describes structural and thermodynamical features together with 
 the inertial dynamics of liquids \cite{marconi2009kinetic,marconi2010dynamic}.

DDFT derives a closed evolution equation for the particle density, $n$, by considering
the local conservation law  for the number of particles 
and relating the  current density 
to the gradient of the density itself by using the rules governing the microscopic
dynamics of the system or some phenomenological argument. The existence of a simple 
relation, useful in practical applications,
between current and density is not a priori guaranteed
and in some cases, depending on the nature of the microscopic dynamics,
it might be more convenient to explicitly 
consider, in addition
to the equation for $n$,  the balance equations for 
the momentum density and energy density variables.
The use of this larger set of variables to describe the state of the system is characteristic of
hydrodynamics or of the kinetic approach, based on the knowledge of the phase space distribution function.

As far as hydrodynamics is concerned
the application of the Navier-Stokes equation at the nanoscale has been questioned, because the presence of
surfaces or the confinement of the particles in very tiny regions may affect drastically  the local relations between
hydrodynamic variables.
Kinetic theory is the natural tool to treat dynamical effects at the microscopic level, but requires a 
large effort because one needs to consider the 
phase space $(\rr,\vv)$ instead of the standard coordinate space. However, recent studies have 
shown that it is possible to conjugate the features of the DFT with those of the discrete numerical technique
introduced about two decades ago to solve kinetic equations, namely the Lattice Boltzmann method (LBM).
The standard LBM  suffers from a lack of physical realism in the treatment
of the interactions and of thermodynamic consistency. It is not fully  microscopically
motivated, but rather is inspired by a top-down approach.
It describes hydrodynamic effects, but has a trivial or "ad hoc" thermodynamics 
corresponding to structureless fluids. 
Shan and Chen \cite{shan1994simulation} and He et al. \cite{he1998discrete} proposed  independently modifications of the LBM aimed to include a better thermodynamic
treatment of the fluids, however none of these approaches were able to capture the 
structural features of fluids which are among the merits of the DDFT theory.

The present paper is organized as follows in sect. \ref{kinetic} we give a short account of the
derivation of the kinetic equation for the single particle phase space distribution function, in section
\ref{overdamped}   we consider the case of a colloidal solution where there is a scale separation between the colloidal particles and the solvent particles
and is possible to represent the system as an assembly of Brownian particles governed by overdamped dynamics. In section \ref{inertial}
we consider instead the case where solute and solvent particles have similar physical properties and the solvent has to be treated explicitly.
The resulting dynamics is inertial  and one has to take into account adequately the hydrodynamic modes.
In section \ref{mixtures} we give explicitly the extension of the kinetic equation to the study of neutral and charged mixtures, while in section
\ref{numerical} we briefly illustrate how the kinetic equation can be
solved numerically using tools borrowed form the lattice Boltzmann method.
Finally in section \ref{conclusions} we present a brief summary.


\section{Kinetic description of inhomogeneous fluids}
\label{kinetic}

Let us consider  an assembly of   
$N$  particles, of mass $m$ and positions $\rr_i$
and velocities $\vv_i$, mutually interacting  through pair potential $U(\rr_i-\rr_j)$ and 
subject to external forces  $\fvec_{ext}(\rr_i)$. The system evolves according to the following set of equations:
\begin{eqnarray}
&\frac{d\rr_i}{dt} &= \vv_i  \nonumber\\
& m \frac{d \vv_i}{dt} & = 
-\bnabri V(\rr_i) 
- \sum_{j(\neq i)} \bnabri U(|\rr_i-\rr_j|) 
- m \gamma \vv_i   + \bfxi_i(t)
\label{kramers-b}
\end{eqnarray}
where the last two terms in \eqref{kramers-b} represent the coupling to a stochastic heat bath at temperature $T$,
characterized by a stochastic white noise forcing $\bfxi_i$  
whose noise amplitude is related to the friction constant, $\gamma$,
by the Einstein fluctuation-dissipation relation
$
\langle \xi^\alpha_{i}(t)
\xi^\beta_{j}(s) \rangle  = 2 \gamma m k_B T\delta_{ij}
\delta^{\alpha\beta} \delta(t-s) ,
$ where $k_B$ is the Boltzmann constant. The noise
represents the effect of microscopic degrees of freedom not 
explicitly accounted for in the interaction terms.

For our purposes it is  more convenient to switch from the coupled set of stochastic
differential equations  \eqref{kramers-b} to the description
based on the N-particle probability density distribution $f_N(\{\rr,\vv\},t)$, where $\{\rr,\vv\}$ indicates a $6N$ dimensional phase space point, which evolves according to 
the Kramers-Fokker-Planck \cite{risken1984fokker}: 
\bea
&& \frac{\partial}{\partial t} f_N(\{\rr,\vv\},t) +\sum_i \Bigl[\vv_{i}\cdot\bnabri-\Bigl(\frac{\
\bnabri V(\rr_i)}{m}+\frac{1}{m} \sum_{j (\neq i)}
\bnabri U(|\xx_i-\xx_j|) \Bigl) \cdot\frac{\partial}{\partial \vv_i} \Bigr]
 f_N(\{\rr,\vv\},t) 
\nonumber \\
&=& 
\gamma\sum_i \Bigr[ \frac{\partial}{\partial \vv_i}  \cdot \vv_i +\frac{k_B T}{m}\frac{\partial^2}{\partial \vv_i^2}\Bigr] f_N(\{\rr,\vv\},t) 
\label{many4}
\eea

The probabilistic
description represented by eq. \eqref{many4} is the result of an ensemble averaging of the
trajectories over a noise ensemble and
over initial conditions in the case of damped stochastic
dynamics ($\gamma>0$) typical of a colloidal suspension. In the case
 Hamiltonian dynamics characteristic of an
atomic liquid, where $\gamma=0$, eq. \eqref{many4} reduces to
the Liouville equation.

The information contained in $f_N$ is fully microscopic and
describes the probability density of the  microstates of the system. 
However, the distribution $f_N$ is a very complicated object to handle 
and provides a representation  redundant for practical purposes. One then
seeks a contracted  description from the $6N$ dimensional $\Gamma$-phase space to
a  $6$ $\mu$-dimensional space, that is, from the phase space distribution of $N$
particles to the one particle phase space distribution,
$f(\rr,\vv,t)$. 
In order to contract the description from the $\Gamma$ to the $\mu$-space, one  considers the
marginalized s-particle distribution functions, $f_s$,
constructed by integrating over the $2 d$ degrees of freedom of  $(N-s)$ particles 
\be
f_s(\rr_1,..\rr_s,\vv_1,..,\vv_s,t)
=\frac{N!}{(N-s)!}\int \prod_{n=s+1}^N d\rr_n d\vv_n \,  f_N (\rr_1,..,\rr_s,\rr_{s+1}..,\rr_N, 
\vv_1,..,\vv_s,\vv_{s+1}..,\vv_N,t).
\label{h1}
\ee
It is straightforward to derive  a set of of coupled  equations connecting the
distribution functions of different orders $s$ among themselves, the so called  Bogoliubov-Born-Green-Kirkwood-Yvon  (BBGKY) hierarchy \cite{hansen1990theory}.
The one particle distribution is $f=f_1$,  which is the object of
the kinetic theory,  is a useful bridge between the many-body molecular dynamics models and 
the continuum mechanics models.
The evolution equation for $f$  , the so called kinetic equation,
contains a non linear term, named collision term, which takes into account the effect of interactions in an approximate fashion.


At the bottom end of the hierarchy, that is
integrating equation~(\eqref{many4}) over $(N-1)$ particle's coordinates and
velocities, one obtains the exact   
evolution equation for the one particle distribution:
\be
\bigl( \frac{\partial}{\partial t}  +\vv\cdot\NN 
-\frac{\nabla V(\rr)}{m}\cdot\frac{\partial}{\partial \vv}\bigr) f(\rr,\vv,t)
= \Omega(\rr,\vv,t)+{\cal B}(\rr,\vv,t).
\label{equilproc}
\ee
Equation  \eqref{equilproc} contains the streaming terms in the l.h.s.,
and the interparticle interaction term, $\Omega$,  the r.h.s. and
a coupling to the stochastic heat bath :
\be
{\cal B}(\rr,\vv,t)=\gamma \Bigl[\frac{k_B T}{m}\frac{\partial^2} 
{\partial \vv^2} 
+\frac{\partial} {\partial \vv}\cdot \vv \Bigr]f(\rr,\vv,t).
\label{lfp}
\ee
For continuous  pair potentials, $U$, one can write:
\be
\Omega(\rr,\vv,t)=
\frac{1}{m}  \frac{\partial}{\partial \vv}  \cdot \int d\rr'\int d\vv' 
f_2(\rr,\vv,\rr',\vv',t){\bf \NN_{\bf r}}U(|\rr-\rr'|) .
\label{bbgky}
\ee
Instead, in the case where $U$ contains an hard-sphere contribution, $\NN_{\bf r}U$ is singular and one has to treat $\Omega$ differently:
the non singular long range part of the potential can still be handled as in  eq.\eqref{bbgky}, while
the  singular piece has to be dealt by special methods of kinetic theory \cite{dufty2005hard,brilliantov2010kinetic}.
In fact, the dynamics of hard spheres is no longer described in terms of forces
since the  trajectories of the particles abruptly change when a pair comes into contact and exchanges
momentum according to the law of elastic collisions. By using the so-called pseudo-Liouville operator approach
one reformulates the dynamics in terms of  binary collision operators rather than forces.
 This method leads to the revised Boltzmann-Enskog kinetic equation describing a system of dense hard-spheres
 and was introduced rigorously
by  Ernst and van Bejeren \cite{van1973modified,van1973modified2} about 40 years ago. In the 
following we shall use this approach to treat the collision term, 
whose explicit form is given  by eq. \eqref{omegaret},
when hard sphere interactions are involved.

In a nutshell, the underlying assumption is that the N-particle distribution function $f_N$ at
all times takes exactly into account the non overlap of any pair of spheres.  Instead, the velocity dependence
of  $f_N$ is factorized into a product of single-particle velocity distribution functions and
the  resulting kinetic equation for the one-particle distribution function, $f(\rr,\vv,t)$, involves the two-particle
distribution function, $f_2$. As suggested by Enskog, $f_2$ 
is  expressed as the product of one particle distribution functions times 
 the positional  pair correlation function $g(\rr,\rr',t)$:
\be
f_2(\rr,\vv,\rr',\vv',t)\approx 
f(\rr,\vv,t) f(\rr',\vv',t) g(\rr,\rr',t)
\nonumber 
\label{factor}
\ee
The approximation employing the product of two one-particle distribution functions is called 
"molecular chaos" hypothesis and it disregards the correlations of the velocities of the two colliding particles, prior to the collision.
Importantly, it decouples the evolution
of $f(\rr,\vv,t)$ from the evolution of the higher order multiparticle distribution 
functions.
Many-particle correlations are however retained  through the structural information contained in
the positional  pair correlation function function $g(\rr,\rr')$. As an approximation, we shall assume that
$g(\rr,\rr')$ is a non local function of the profile $n(\rr,t)$, depends on time only
through the density profile and has the same form as in a nonuniform 
equilibrium state whose density is $n(\rr,t)$. 

In order to describe the properties of molecular fluids we need
the hydrodynamic framework to consider five field variables as the minimal ingredients to describe
the local (macro)state of a simple fluid. These fields  are the  first velocity moments of
$f(\rr,\vv,t)$ and correspond to the
number density, local velocity
and  local temperature of the fluid:
\be
\left( \begin{array}{ccc}
n(\rr,t)  \\
 n(\rr,t)\uu(\rr,t)\\
 \frac{3}{2} k_B n(\rr,t)T(\rr,t)\\  \end{array} \right) =
\int d\vv
\left( \begin{array}{ccc}
1   \\
 \vv   \\
\frac{m(\vv-\uu(\rr,t))^2}{2}  \end{array} \right)  f(\rr,\vv,t).
\label{colonna}
\ee
The hydrodynamic fields  obey the following balance equations, obtained by multiplying 
the kinetic equation by the first Hermite tensorial polynomials, proportional to  $\{1,\vv,(\vv-\uu(\rr,t))^2\}$, and integrating over the velocity :
\bea
&&\partial_{t}n(\rr,t) +\nabla\cdot ( n(\rr,t) \uu(\rr,t))=0
\label{continuity}
\\
&&m n(\rr,t)[\partial_{t}u_j(\rr,t)+u_i(\rr,t) \nabla_i u_j(\rr,t)] 
+\nabla_i P_{ij}^{(K)}(\rr,t) +n(\rr,t) \nabla_j V(\rr)-n(\rr,t) \Phi_j(\rr,t)=
b_j^{(1)}(\rr,t)
\label{momentum}\\
&&\frac{3}{2}k_B n(\rr,t) [\partial_{t} +u_i(\rr,t)\nabla_i] T(\rr,t)+P^{(K)}_{ij}(\rr,t) \nabla_i u_j(\rr,t)
+\nabla_i q^{(K)}_i(\rr,t)-Q (\rr,t) =b^{(2)}(\rr,t) .
\nonumber\\
\label{energy}
\eea
We have introduced the kinetic component of the pressure
tensor
$P_{ij}^{(K)}$  and the  heat flux  vector $q^{(K)}_{i}$,
\be
\left( \begin{array}{ccc}
P^{(K)}_{ij}(\rr,t)  \\
q^{(K)}_{i}(\rr,t)\\
 \end{array} \right) =
\int d\vv
\left( \begin{array}{ccc}
  (\vv-\uu)_i(\vv-\uu)_j  \\
 \frac{m}{2} (\vv-\uu)^2(\vv-\uu)_i   \end{array} \right)  f(\rr,\vv,t).
\label{colonna}
\ee
 The 'source' terms stem from the heat bath
\be
\left( \begin{array}{ccc}
b_i^{(1)}(\rr,t)  \\
b^{(2)}(\rr,t) \\
 \end{array} \right) =
\int d\vv
\left( \begin{array}{ccc}
  (v-u)_i  {\cal B}(\rr,\vv,t)\\
  \frac{m}{2}
  (\vv-\uu)^2 {\cal B}(\rr,\vv,t)\\
  \end{array} \right)  = -\gamma m n(\rr,t)
\left(  \begin{array}{ccc}
u_i(\rr,t)\\
<\vv^2>-3 \frac{k_B }{m}T(\rr,t)
  \end{array} \right)
  \label{colonna}
\ee
Finally, one considers the terms due to  interparticle interactions:
\be
n(\rr,t)\Phi_i(\rr,t)  =m\int d\vv   \Omega(\rr,\vv,t) v_i
= -\nabla_j  P_{ij}^{(C)}(\rr,t)
\label{c1}
\ee
It can be shown that $\Phi_i$ is related
to a
spatial derivative of the
potential part of the stress tensor $P_{ij}^{(C)}$ while
\be
Q (\rr,t) =\frac{m}{2}
\int d\vv  \Omega(\rr,\vv,t)  (\vv-\uu)^2
\label{c2}
\ee
 is a combination of the gradient of the collisional contribution
to the heat flux vector $q_{i}^{c}$ and the pressure tensor times
the strain rate: 
\be
Q(\rr,t)= -\nabla_{j}q_{j}^{c}(\rr,t)-P_{ji}^{c}(\rr,t)\nabla_{j}u_{i}(\rr,t).\nonumber \\
\label{gradienth}
\ee
Both $\Phi_i$ and $Q$
 vanish in uniform systems.

The difficulty with eqs. \eqref{continuity}-\eqref{energy} is not only due to the non linearity
stemming from the interaction terms, but also from the 
projection of  the kinetic equation onto the velocity space ( spanned by the various Hermite tensorial polynomials ).
In fact, the procedure eliminates the velocity dependence, but on the other hands
generates an infinite hierarchy of
equations for the evolution of the velocity moments of
$f(\rr,\vv,t)$
due to the coupling determined by the presence of the convective term $\vv\cdot \nabla_{\rr} f$.
Notice that $P^{(K)}_{ij}$ and $q^{(K)}_{i}$
cannot in general be expressed in terms of the five hydrodynamic moments.
To obtain a closure, one must truncate this
hierarchy at a given level, which requires the approximation of
higher moments. 
For example, for the standard truncation at the velocity
($n=1$) level (i.e.\ the same level of description as the
Navier-Stokes equations), one must control terms (in appropriate
units) of the form $\int d \vv \, (\vv \otimes \vv -I)
f(\rr,\vv,t)$, where $I$ is the 3$\times$3 identity matrix.
One way to deal with such a problem is to introduce phenomenologically motivated
relations which allow to express higher order velocity moments in terms of the
five basic moments, these are the so called
constitutive equations for the momentum and heat flux.
In the case of systems described by
 (under)damped dynamics it is possible to eliminate all moments higher than the zeroth moment, 
 in favor of $n(\rr,t)$ by  multiple time scale  expansion  \cite{marconi2007phase}.
 
 However, as recently proposed by our group it is possible to solve  directly the kinetic equation
 even in the presence of non trivial collision terms by using the 
 Weighted Density Lattice Boltzmann Method  (WDLBM)  which combines the structural features of the DDFT
 with the requirements of hydrodynamics and also
gives transport coefficients in self-consistent fashion, see section \ref{inertial}.

To proceed further we must give explicitly the form of $\Omega$ and discuss how to treat it.
In order to obtain a suitable description of the fluid behavior,  reproducing at least the qualitative feature of the
equation of state and the density dependence of the transport coefficients the
interaction must be separated into short range repulsion (assimilable to hard spheres) and long range 
attractive contributions:
\be
{\Omega}(\rr,\vv,t)=
{\Omega}_{rep}(\rr,\vv,t)+{\Omega}_{attr}(\rr,\vv,t)\nonumber
\ee
A good choice is to assume for
${\Omega}_{rep}$ the Revised Enskog Theory (RET)  form for hard spheres  \cite{van1973modified,van1973modified2}:
\bea
&&{\Omega}_{rep}(\rr,\vv,t)
= \sigma^{d-1}\int d\vv_2\int 
d\hat{\ss}\Theta(\hat{\ss}\cdot \vv_{12}) (\hat{\ss} 
\cdot \vv_{12})\times\nonumber\\
&&\{ g_2(\rr,\rr-\ss\shat) f (\rr,\vv')f (\rr-\ss\shat,\vv_2')  -
g_2(\rr,\rr+\ss\shat)f(\rr,\vv)f(\rr+\ss\shat,\vv_2)\}
\label{omegaret}
\eea
where the primes indicate post-collisional velocities
$\vv'=\vv-(\hat\ss\cdot\vv_{12})\hat\ss\,\,\,$ and  $\,\,\,\vv_2'=\vv_2+(\hat\ss\cdot\vv_{12})\hat\ss$.
determined by the conservation of total momentum and total energy in an elastic collision; $\sigma$
is the hard-sphere diameter and $\shat$ the unit vector directed from the center of sphere $1$ to 
the center of sphere $2$.
Interactions are non-local
 and momentum and energy can be transferred instantaneously across finite distances when the spheres collide 
An adiabatic  approximation for $g_2(\rr,\rr',t)$ is chosen by which
 $g_2(\rr,\rr',t)$ is assumed to be given by the equilibrium pair correlation function of the system
 when its ensemble averaged density is given by $n(\rr,t)$.
 In addition since the exact form of the inhomogeneous pair correlation $g_2$ is not known, one uses the
Fischer-Methfessel prescription to construct it locally \cite{fischer1980born}.
This ansatz is common both to the DDFT and the WDLBM methods.
 The calculation of the collision integral involves
the  inhomogeneous pair correlation function, $g_2$,  at contact which is  
not known. One resorts to 
the Fischer and Methfessel prescription\cite{fischer1980born}, which 
 assumes that the functional form of the  inhomogeneous  $g_2(\rr,\rr+\sigma\bk)$ is obtained by
replacing the  density $n(\rr)$ by the associated coarse grained  density
$\bar n (\rr)$. The latter  is the average of  $n(\rr)$
over a sphere of diameter $\sigma$ centered at $\rr$:
$$
\bar n(\rr)=\frac{1}{ \frac{\pi}{6} \sigma^3
  }\int d\rr'
n(\rr+\rr')\, \theta( \frac{\sigma}{2}-|\rr-\rr'|)
$$
The following rule gives the pair correlation function at contact:
\be
g_2(\rr,\rr+\sigma \bk)
=g_{2}^{bulk}(\bar\eta(\rr+\frac{1}{2}{\bk} \sigma))
\ee
where the local average packing fraction $\bar \eta$ is
$
\bar\eta(\rr) =\frac{\pi}{6} \bar n
(\rr) \sigma^3
$
and the explicit expression of $g_{2}^{bulk}$ is provided by
the  Carnahan-Starling equation \cite{hansen1990theory}:
\be
g_{2}^{bulk}(\eta)
=
\frac{1-\eta/2}{(1-\eta)^3}.
\label{carnahan}
\ee

Finally, the  attractive contribution to the collision term is written in the mean-field form
\be
{\Omega}_{attr}(\rr,\vv,t)= 
\frac{1}{m}{\NN_{\bf v}} \cdot \int d\rr'\int d\vv' 
f(\rr,\vv,t) f(\rr',\vv',t){\bf \NN_{\bf r}}U(|\rr-\rr'|) 
\ee
where configurational correlation are disregarded having set $g(\rr,\rr',t)=1$ in this formula.
Due to such crude approximation the attractive term contributes to the equation of state,
but does change the transport coefficients, with the notable exception of the
diffusion coefficient. 

\section{Overdamped dynamics}
\label{overdamped}
In this section we shall specialize to the over damped  dynamics and obtain the DDFT equation of evolution.
Classical density functional theory (DFT) has been used with great
success to investigate the structural  and thermodynamic properties 
of inhomogeneous 
classical fluids~\cite{evans1979nature}. It 
represents a relevant generalization of the highly successful and rigorous
method originally introduced about 50 years ago by Hohenberg and Kohn
in quantum many body theory.
The widespread use of the DFT approach is due to the fact that
its fundamental entity, the Helmholtz free energy, $F[n]$, is an 
intrinsic functional  of the local density of molecules $n(r)$, i.e. is
independent of the external potential to which the fluid is subjected, 
in other words   $F$ is a universal functional once the interactions among the particles 
and their properties are
specified. The theory  states that for a fixed external
potential $V(r)$, fixed temperature and chemical potential, 
there is a unique equilibrium density
profile $n(r)$, which can be found by minimizing the Grand potential
functional $\Omega[n]=F[n]+\int n(r)[V(r)-\mu]dr$ with respect to $n$.
Although
$F[n]$ is known exactly only in few particular cases,
fairly good approximations have been devised, so that the method is
versatile and generally applicable with success to 
study the properties of real systems under a
variety of thermodynamic and geometric conditions.
A variety of problems has been solved,
ranging from  adsorption, phase transitions at
surfaces, wetting phenomena, fluids confined in narrow
pores, theory of freezing, depletion forces, etc.  

Under appropriate conditions such as those realized in colloidal
solutions, recent studies have shown that a dynamical extension of the DFT,
the so called dynamical density functional theory (DDFT)   
may provide a valid description of the main features of the approach towards 
equilibrium. 
In order to establish the DDFT equation of evolution within the present framework
one must consider the role of the two  terms $b_i^{(1)}, b^{(2)}$, 
in  eqs. \eqref{momentum}-\eqref{energy}, accounting for the
drag force proportional to the particle velocity and the random
stochastic acceleration due to the solvent atoms impinging on the molecules.
These terms determine a fast  equilibration process of the momentum current
$n\uu$
and of the local temperature $T$ towards their stationary
values when
the friction $\gamma$  is large
(overdamped limit). In this case,  one may  neglect the inertial term $d \vv_i/dt$  in eq.  \eqref{kramers-b}
and derive the N-particle Smoluchowski equation associated with such first order 
system, a time dependent partial differential equation for 
the distribution function of the $N$ particle positions. 
This is the starting point of the strategy followed by Archer and Evans \cite{archer2004dynamical}
to derive their version of the DDFT by integrating out the coordinates
of $(N-1)$ particles.
The presence of the friction terms is of great help  in reducing the infinite hierarchy, 
 whose
five equations appearing in eqs. \eqref{continuity}-\eqref{energy} are only the first members,
to the single equation for the density  appearing in the DDFT equation. 

The rigorous mathematical procedure to derive
this result employs the
multiple time scale analysis \cite{marconi2006nonequilibrium,marini2007theory},
which exploits the
time scale separation between
the zeroth mode associated with the density fluctuations 
and the remaining modes, which also include the momentum and energy
fluctuations. Due to the action of the heat bath  these are fast relaxing modes
because the velocities of the particles rapidly relax towards the
equilibrium distribution, in a  time of order
of the inverse friction time $\tau=1/\gamma$. 
Since the momentum and energy of the colloidal particles are not conserved,
their currents become rapidly
``slaved'' to the density, i.e. the 
evolution is completely determined in terms of $n(\rr,t)$ and its derivatives.
The only relevant evolution on  timescales larger than $\tau=1/\gamma$ regards
the spatial distribution of the particles. 
 This is the
reason why the DDFT gives a sufficiently accurate
description of colloidal systems.
 One can determine
the particle current, ${\bf J}(\rr,t)=n(\rr,t) \uu(\rr,t)$,  by imposing the vanishing 
of the inertial terms in  eq. \eqref{momentum}
and using eq. \eqref{c1}, so that from the momentum balance the following approximate equality 
holds:
\be
\nabla_i P_{ij}^{(K)}(\rr,t) + \nabla_i P_{ij}^{(C)}(\rr,t)  +n(\rr,t)\nabla_j V(\rr) \approx
- m \gamma n(\rr,t) u_j(\rr,t) .
\ee
Finally,
substituting $n u_j$  into  the continuity equation \eqref{continuity} one finds:
\be
 \frac{\partial{n} (\rr,t)}{\partial t} =\frac{1}{m \gamma}\sum_i
\nabla_i \Bigl[ \nabla_j  P_{ij}^{(K)}(\rr,t)  +\nabla_j  P_{ij}^{(C)}(\rr,t)   +n(\rr,t)\nabla_i V(\rr,t)\Bigr].
\label{DDFTeq}
\ee
Neglecting dissipative contributions 
(viscous and thermal conduction, see eq.  \eqref{splitphi} below) 
to the pressure tensor we can write $P_{ij}^{(K)}(\rr,t) =k_B T n(\rr,t) \delta_{ij}$
or equivalently
$$
 \nabla_j  P_{ij}^{(K)}(\rr,t)=n(\rr,t)\nabla_i  \frac{\delta  {\cal F}_{ideal}[{n}]}{\delta {n}(\rr,t)} $$ 
 where
 $${\cal F}_{ideal}[n]= 
k_B T \int d\rr n(\rr,t)(\ln n(\rr,t)-1).$$
is the ideal gas contribution to the free energy.

Similarly one finds the following relation between the excess pressure and the
excess free energy, ${\cal F}_{int}$ (see  eq. \eqref{meanforce} below and references):
 $$ \nabla_j  P_{ij}^{(C)}(\rr,t)=  n(\rr,t)\nabla_i  \frac{\delta  {\cal F}_{int}[{n}]}
{\delta {n}(\rr,t)} $$ 
Collecting together one obtains the following equation
\be
\frac{\partial{n} (\rr,t)}{\partial t}=\frac{1}{m \gamma}
\nabla_i  n(\rr,t)\left[ \nabla_i
\left(\frac{\delta {\cal F}_{ideal}[{n}]}
{\delta {n}(\rr,t)}
+ \frac{\delta  {\cal F}_{int}[{n}]}
{\delta {n}(\rr,t)} + V(\rr)  \right)\right].
\label{due}
\ee

One can recognize that  equation \eqref{due} for the density corresponds to 
the DDFT equation. The details of the derivation of the DDFT formula  can be found
in ref. \cite{marconi1999dynamic}. 
 Such a formula has a kinetic derivation, but one can make contact with the
equilibrium density functional theory by  considering
the evolution of the statistical average of the
microscopic instantaneous density over an ensemble
of identical copies of the original system, each copy being
characterized by a different realization of the noise.  Such a
procedure, besides being easily realizable in a numerical simulation,  
leads to a simplification of the equation \eqref{due} and to
the  interpretation of the drift term as the functional gradient of the derivative of the free energy functional
with respect to the density, that is the local chemical potential gradient.

The crucial assumptions made in deriving eq. \eqref{due}  are the following:
a) the density $n(\rr,t)$ is the average of the microscopic instantaneous 
density over the realizations of the random noise and is therefore a
smooth function of the coordinate $\rr$. 
b) The functional $F[n]$ is a function solely of the density $n(\rr,t)$.
c) The instantaneous two-particle correlations are 
contained in ${\cal F}_{int}[n]$ and are approximated 
by those of an equilibrium system having  the same density profile
as the system at time $t$.  This is the so called adiabatic approximation.
Since in DDFT the driving force towards the steady state is given by a derivative of the free energy
it cannot contain contributions from
dissipative forces. Such an aspect is at variance with the kinetic approach where, as shown below,
frictional and viscous forces appear naturally in the stress tensor.
From eq. \eqref{due}
one can derive an H-theorem  
stating that the free energy  never increases during the relaxation process.

Notice that the theory has selected a particular reference frame, where the solvent is at rest.
Such a choice breaks the translational invariance of the system, which for this reason cannot support sound waves. 

The DDFT has been applied to several problems and extended to treat non simple fluids
and more complicated interactions such as the Hydrodynamic interaction \cite{goddard2012general}.
It can be extended to describe charged and uncharged fluid mixtures and applies whenever the 
system dynamics is a diffusion relaxation process. Therefore it can be used to describe
the dynamics of ions in a aqueous solvent, if the structure of the latter is not important 
for the problem.

The overdamped dynamics illustrated above is the extreme limit where $\gamma \sigma>> v_T$, where
$v_T$ is the thermal velocity. However, one can investigate corrections for finite values of $\gamma$ 
\cite{wilemski1976derivation,titulaer1978systematic}.
These corrections can be computed by an inverse friction expansion or by multiple time scale
methods and give rise to a more complex DDFT equation as shown in refs. \cite{marconi2006nonequilibrium,marini2007theory} , which takes into account 
momentum and energy fluctuations.

\section{Inertial dynamics}
\label{inertial}
The inverse friction expansion does not
help in the case of molecular fluids where the inverse friction parameter
$1/\gamma$ diverges, and only internal dissipation mechanisms are at work.
We consider now the case with the $\gamma=0$ ( $b^{(i)}=0$ ).
A salient feature of molecular liquids  is their ability to  support
hydrodynamical modes, since  
 particle number,  momentum and energy are locally conserved.
The  set of evolution equations
for the five hydrodynamic variables do not form a closed
system, unless one assumes some phenomenological constitutive
relations between the gradients of $ P_{ij}$ and ${\bf q}$ (the sums of
the kinetic and collisional parts) and the fluxes. 

Using  the kinetic equation it is possible to approximate the non linear interaction term 
and obtain its explicit dependence on the small set of
hydrodynamic modes and not on the full distribution function $f(\rr,\vv,t)$
 obtaining a fast numerical solution of the equation. Such a step is very important in our treatment because
 it allows to reduce enormously the computational effort and obtain a fast numerical solution of the kinetic equation.
 Such a reduction must
preserve the translational invariance of the system, i.e. must
not select the reference frame where the solvent is at rest \cite{anero2007dynamic,anero2013functional,archer2009dynamical}.

A practical treatment of the RET collision operator was suggested by Dufty et \emph{al} 
\cite{dufty1996practical,santos1998kinetic}, who proposed to separate the contributions to $\Omega$
stemming from  the hydrodynamic modes from those from the
non-hydrodynamic modes. Such a goal is achieved by projecting the collision term onto the
hydrodynamic subspace spanned by the functions $\{1,\vv,v^{2}\}$
and onto the complementary kinetic subspace: 
\begin{equation}
\Omega={\cal P}_{hydro}\Omega+(I-{\cal P}_{hydro})\Omega\label{splitting}
\end{equation}
 with 
\begin{equation}
{\cal P}_{hydro}\Omega=\frac{1}{k_{B}T(\rr,t)}\phi_{M}(\rr,t)\Bigl[(\vv-\uu)\cdot n(\rr,t) \Phi(\rr,t)+[\frac{m(\vv-\uu(\rr,t))^{2}}{3k_{B}T(\rr,t)}-1]Q(\rr,t)\Bigr] 
\end{equation}
where $\Phi$ and $Q$ are given by eqs. \eqref{c1} and \eqref{c2}, respectively and
\[
\phi_{M}(\rr,\vv,t)=[\frac{m}{2\pi k_{B}T(\rr,t)}]^{3/2}\exp\left(-\frac{m(\vv-\uu)^{2}}{2k_{B}T(\rr,t)}\right)
\]
is the local Maxwellian. 
As far as the projection of $\Omega$ onto the non-hydrodynamics sub-space
Dufty and coworkers approximated it by a  phenomenological  Bathnagar-Gross-Krook (BGK) \cite{bhatnagar1954model} prescription, which    preserves the
number of particles, the momentum and the kinetic energy, thus
fulfilling the physical symmetries and conservation laws of
the fluid:
\begin{equation}
(I-{\cal P}_{hydro})\Omega\approx-\nu \Bigl[f(\rr,\vv,t)-n(\rr,t)\phi_{M}(\rr,\vv,t)\Bigr]  \label{dufty1}
\end{equation}
This part of the collision operator contributes to determine the values
of the transport coefficients which turn out to be functions of phenomenological
parameter $\nu$, a phenomenological collision frequency chosen as to reproduce
the kinetic contribution to the viscosity.
By inserting the approximation above and neglecting the 
temperature gradients we rewrite the evolution equation for $f(\rr,\vv,t)$ as:
\bea
\bigl( \frac{\partial}{\partial t}  +\vv\cdot\NN 
-\frac{\nabla V(\rr)}{m}\cdot\frac{\partial}{\partial \vv}\bigr) f(\rr,\vv,t)-\frac{{\bm\Phi}(\rr,t) }{k_B T} \cdot(\vv-\uu(\rr,t))  n(\rr,t)\phi_M(\rr,\vv,t)= -\nu[f(\rr,\vv,t)- n(\rr,t)\phi_M(\rr,\vv,t) \nonumber]\\
\label{equilproc2}
\eea  
It is interesting to analyze the structure of the self consistent field which appears in the equation of evolution
and is obtained by using \eqref{omegaret} in \eqref{c1}.
One finds that the field ${\bm \Phi}$ is given by the sum of different forces, dissipative and non dissipative:
\be
{\bm \Phi}(\rr,t)= \FF^{mf}(\rr,t)+\FF^{visc}(\rr,t)+\FF^{T}(\rr,t)  .
\label{splitphi}
\ee
where the first term represents the gradient of the potential of mean force, that is the average force of the remaining particles
on a particle at $\rr$, the second term describes the viscous force due to the presence of velocity gradients, whereas 
the last one  is due to the existence of thermal gradients.
All terms can be identified and computed self-consistently  once we know the fields $n,\uu,T$.
Only the first force is non vanishing at equilibrium. For such a reason in the DDFT, which only considers
the equilibrium free energy, the effective force is purely non dissipative.
   In fact $\FF^{mf}(\rr,t)$  may only describe forces which have an equilibrium counterpart
entropic, depletion, electrostatic, Van der Waals, etc, but not velocity dependent forces or forces due to thermal gradients.

Using the explicit form of the RET collision \eqref{omegaret}  the following expressions for the
different forces are derived:
\begin{equation}
F_{i}^{mf}(\rr,t)=-k_{B}T\sigma^{2}\int d\hat{k}k_{i}g_{2}(\rr,\rr+\sigma\hat k,t)n(\rr+\sigma\hat k,t).
\label{meanforce}
\end{equation}
where one can show that for small density  gradients: 
$\,\,
\FF^{mf}(\rr,t)= -\NN \mu_{exc}(\rr,t).$
The expression of the viscous force reads
$$
F_i^{visc}(\rr,t)=
2\sigma^2 \sqrt{\frac{m k_B T}{\pi} }
\int d\hat{k}  k_{i} g_2(\rr,\rr+\sigma\hat k,t)
n(\rr+\sigma\hat k,t)
 k_j
(u_j(\rr+\sigma \hat k,t)-u_j(\rr,t)) 
$$
 and the force due to the presence of thermal gradients is
\begin{equation}
F_{i}^{T}(\rr,t)=-\frac{\sigma^{2}}{2}\int d\hat{k}k_{i}g_{2}(\rr,\rr+\sigma \hat k,t)n(\rr+\sigma\hat k,t)k_{B}[T(\rr+\sigma\hat k,t)-T(\rr,t)].\label{thermalforce}
\end{equation}
Notice that  all these forces are expressed as convolutions 
involving $g_2(\rr,\rr',t), n(\rr,t), \uu(\rr,t), T(\rr,t)$ .
From the equations above one can derive the following explicit expressions for the collisional contributions to the bulk
dynamical  viscosity coefficient:
\be
\eta^{(C)}=\frac{4}{15} \sqrt{\pi m k_B T}\, \sigma^4 g_2(\sigma) n^2 
\ee
and to the heat conductivity:
$$\lambda^{(C)}=\frac{2}{3}\frac{k_{B}}{m}\sqrt{m\pi k_{B}T}g_{2}(\sigma)n^{2}\sigma^{4}$$
which correspond to the formulae proposed by Longuet-Higgins and Pople \cite{longuet1956transport}.

Higher order corrections which better describe the density dependence of transport coefficient
 have also been included by employing an extended approximation which takes into account
 the dependence of the relaxation time $\nu$ on the hydrodynamic modes
\cite{marconi2013weighted}.

Before closing this section we would like to mention the important work performed in a spirit similar
to our work by L.S. Luo and coworkers \cite{luo2011numerics,luo2010lattice} and more recently by 
Baskaran and Lowengrub \cite{baskaran2013kinetic}.

\section{Applications of the WDLBM to neutral and  charged mixtures}
\label{mixtures}

In the present section we briefly discuss  recent extensions and applications of the
 self-consistent dynamical method,  to binary neutral    or to
ternary charged mixtures comprised of
positively and negatively charged hard spheres carrying point like 
charges (the ions) plus
neutral spheres representing the solvent \cite{oleksy2009microscopic}.
 Each species, denoted by $\alpha$, has mass $m^\alpha$, diameter $\sigma^{\alpha\alpha}$.
 In the case of the binary mixture there are no charges (valence $z^\alpha=0$) and there are only two components,
 while for the Coulomb case the two ionic components carry  
 charges $z^\alpha e$, $e$ being the proton charge .
 In the latter case the index $\alpha=0$, identifies the solvent ,
whose valence is zero, while
$\alpha=\pm$ identifies the two oppositely charged ionic species interacting through a uniform medium
 of  constant dielectric permittivity. 

We describe the system by the following set of  Enskog-like equations   governing
the evolution 
one-particle phase space distributions $f^\alpha(\rr,\vv,t)$ of each species \cite{marini2012charge}:
\begin{align}
& \frac{\partial}{\partial t} f^\alpha(\rr,\vv,t) +\vv\cdot\NN f^\alpha (\rr,\vv,t)
+\frac{\FF^{\alpha}(\rr)}{\ma}\cdot
\frac{\partial}{\partial \vv} f^\alpha (\rr,\vv,t)
=\nonumber\\
& -\nu[f^\alpha(\rr,\vv,t)- \na(\rr,t)\phi^{\alpha}_{\perp}(\rr,\vv,t)]+\frac{{\bm\Phi}^{\alpha}(\rr,t) }{k_B T} \cdot(\vv-\uu(\rr,t)) 
\na(\rr,T) \phi^{\alpha}(\rr,\vv,t) =\frac{e z^\alpha }{m^\alpha}\nabla \psi(\rr)\cdot
\frac{\partial}{\partial \vv} f^\alpha(\rr,\vv,t)
\label{evolution}
\end{align}
The BGK term is modified with respect to eq \eqref{equilproc2} and contains the
distributions functions $\phi^{\alpha}_{\perp}$ and $\phi^{\alpha}$
which have the following representations (see ref. \cite{marconi2011multicomponent} for details):
\be
\phi^{\alpha}(\rr,\vv,t)=[\frac{\ma}{2\pi k_B T}]^{3/2}\exp
\Bigl(-\frac{\ma(\vv-\uu(\rr,t))^2}{2 k_B T} \Bigl)
\label{psia}
\ee
and
\bea
&&
\phi^{\alpha}_{\perp}(\rr,\vv,t)=\phi^{\alpha}(\rr,\vv,t) \Bigl\{1+
\frac{\ma(\uua(\rr,t)-\uu(\rr,t))\cdot(\vv-\uu(\rr,t))}{k_B T}\nonumber\\
&&
+\frac{\ma}{2 k_B T}
\Bigl(\frac{\ma\bigl[(\uua(\rr,t)-\uu(\rr,t))\cdot(\vv-\uu(\rr,t))\bigl]^2}
{k_B T}-\bigl(\uua(\rr,t)-\uu(\rr,t)\bigl)^2\Bigl)\Bigl\}. \nonumber\\
\label{prefactor}
\eea

In the case of a one-component fluid there is no difference between $\phi^{\alpha}_{\perp}$ and $\phi^{\alpha}$,
since the velocities $\uu^\alpha$ and $\uu$ coincide and the standard  BGK approximation
involves the difference between the distribution $f^\alpha$ and the
local  Maxwellian $\na(\rr,t)\phi^\alpha(\rr,\vv,t)$. 
The reason to use the modified distributions   \eqref{prefactor} instead of \eqref{psia}  
is to obtain the correct mutual diffusion and hydrodynamic properties starting from
\eqref{evolution}. By taking a simple BGK recipe, that is, by setting $\phi^{\alpha}_{\perp}=\phi^{\alpha}$,
would lead to a double counting of the interactions on the diffusive properties.

The effective field acting on each species, ${\bm \Phi}^{\alpha}(\rr,t)$, stems from interparticle forces
and can now be separated in the case of isothermal systems into three contributions by generalizing  eq. \eqref{splitphi} \cite{marconi2011non}  as
 \be
{\bm \Phi}^{\alpha}(\rr,t)= \FF^{\alpha,mf}(\rr,t)+\FF^{\alpha,drag}(\rr,t)+\FF^{\alpha,visc}(\rr,t) .
\label{splitforce}
\ee 
The new term $\FF^{\alpha,drag}$ is a frictional force resulting from the fact that the velocities of each species can be different
from the barycentric fluid velocity and depends linearly on their relative velocities:
\be
\FF^{\alpha,drag}(\rr,t)= 
-\sum_\beta {\bm \gamma}^{\alpha\beta} (\rr,t)  (\uu^\alpha(\rr,t)-\uu^\beta(\rr,t))
\label{dragforce}
\ee
where $\gamma_{ij}^{\alpha\beta}$ is the inhomogeneous friction tensor given by:
\be
\gamma_{ij}^{\alpha\beta}(\rr,t)=2(\sab)^2 \sqrt{\frac{2\muab k_B T}{\pi} }
\int d\bk \hat k_i \hat k_j
\gab(\rr,\rr+\sab\bk,t)
n^\beta(\rr+\sab\bk,t),
\label{tensorgamma}
\ee
where  $\muab$ 
is the reduced mass $\muab=\frac{\ma \mb}{\ma+\mb}$, $\sab$
the arithmetic average of the diameters for the colliding pair and $\gab$ is the pair correlation at contact
for species $\alpha$ and $\beta$ whose value is obtained  from the
extension  of the  Carnahan-Starling equation to mixtures \cite{boublik1970hard,mansoori1971equilibrium,wendland1997born}.
We have shown (see ref. \cite{marconi2011non}) that when there exists a disparity of masses and concentrations
between the two species of a binary mixture, the heavier and diluted component can be treated as an assembly
of colloidal particles moving in a fluid assimilated to a solvent, whose effect is to exert a drag force
on the colloidal particles.
One eliminates the fluid variables relative to the light component from the description of the composite system 
and obtain a closed equation for the heavy particles only . To achieve that, one assumes that 
the dynamical properties of the heavy particles evolve on a time-scale much longer than 
the characteristic time-scale of the light particles.
Calling $n^c(\rr,t)$ the number density of colloidal particles and $n^s(\rr,t)$ the analogous quantity for the solvent
particles in the dilute limit,  $n^c/n^s<<1$, one obtains the following 
advection-diffusion equation:
\bea
\partial_{t}n^c(\rr,t)+\nabla\cdot (\uu(\rr,t) n^c(\rr,t))=
\frac{1  }{\gamma}\nabla\cdot \Bigl[ n^c(\rr,t)
\Bigl( \NN \mu^{c}(\rr,t) -\FF^{c}(\rr)\Bigl)\Bigl],
\nonumber\\
\label{ddft}
\eea 
where for a bulk system the friction coefficient,  $\gamma$,  is given by the expression:
\be
\frac{1}{\gamma}=\frac{3}{8 n^s}\frac{1}{ \sqrt {\pi m^s k_B T} \sigma_{cs}^2 g_{cs}(\sigma_{cs}) } 
\label{dab}
\ee
and 
the local chemical potential of the colloidal particles
is determined by the colloidal-colloidal and by the colloidal-solvent interactions, $\mu^c(\rr,t)$,   (see refs. \cite{{marconi2011dynamics,marconi2011multicomponent},marconi2011non} ):
$$
\NN\mu^c(\rr,t)= k_B T\NN \ln n^c(\rr,t)-\FF^{c,mf}(\rr,t),
$$
We, now, observe that eq.\eqref{ddft} 
is formally identical to a  DDFT equation for the $c$ species in a velocity field $\uu(\rr,t)$.
As $n^c\to 0$ the gradient of the chemical potential $\mu^c$  approaches the ideal gas value, $k_B T \NN n^c(\rr,t)$,
so that  eq. \eqref{ddft} becomes a linear advection-diffusion equation for the field $n^c$, with a diffusion coefficient
given by:
\be
D= \frac{k_B T}{\gamma},
\ee
to be interpreted as a fluctuation-dissipation relation between $\gamma$ and $D$ .
 
\subsection{Charged mixtures}

In the case of charged mixture one has to take separately into account 
the electrostatic potential  $\psi(\rr,t)$ generated by the charge distribution $\rho_e(\rr,t)=e(n^+(\rr,t)-n^-(\rr,t))$
 (where $n^\pm$ are the zeroth velocity moments of $f^\pm$)
and by the fixed charges located on the pore surfaces and on the electrodes and satisfies the Poisson equation:
\be
\nabla^2 \psi(\rr,t)= -\frac{\rho_e(\rr,t)}{\epsilon}
\label{Poisson}
\ee
with boundary conditions $-\nabla \psi(\rr,t)\cdot \hat{n}=\Sigma(\rr)/\epsilon$ at the confining surfaces,
where $\Sigma(\rr)$ is the surface charge density sitting on the boundaries and $\hat{n}$ is
the unit vector normal to the surface.

In the limit of slowly varying fields 
the transport equations \eqref{evolution} must  reproduce
the equations of Electrokinetics.
In order to recover the  equation describing the coupling between diffusion and drift induced by the presence 
of electric field, the so called Poisson-Nernst-Planck equation, one uses 
the momentum balance equation for the species $\alpha$ and neglects inertial terms
and finds
\be
\frac{\partial n^\pm(\rr,t)}{\partial t}+{\bm \nabla}\cdot {\bf J}^\pm(\rr,t)=0
\ee
with
 \be
{\bf J}^\pm_i(\rr,t)=  -\frac{1}{\gamma^\pm} n^\pm(\rr,t) \nabla \mu^{\pm}(\rr,t) 
-\frac{1}{\gamma^\pm}   e z^\pm n^\pm(\rr,t)\nabla \psi(\rr,t)  + n^\pm(\rr,t) \uu(\rr,t)
 \label{microcurrent}
\ee
where $\gamma^\pm$  represents the  drag coefficient due to the frictional force exerted by the fluid on the particles of type $\alpha=\pm$,
in reason of their drift velocities
and the average barycentric velocity $\uu$ is given by
\be
\uu(\rr,t)=\frac{\sum_{\alpha}\ma\na(\rr,t)\uua(\rr,t)}
{\sum_{\alpha}\ma\na(\rr,t)}
\ee

In dilute solutions  the
charged components are expected to experience a large friction arising only from the solvent 
while a negligible friction  from the oppositely charged species, so that we further approximate
and the friction can be evaluated in uniform bulk conditions to be
\be
\gamma^{\pm} \approx \frac{8}{3} \sqrt{2 \pi  k_B T  \frac{m^{\pm} m^0}{m^{\pm} + m^0 }}
g^{0\pm}n^0 (\sigma^{0\pm})^2 
\ee
with $n^0$ being the bulk density of the solvent and 
$g^{0\pm}$ the bulk ion-solvent pair correlation function evaluated at contact.
In stationary conditions, the following approximated force balance is obtained
\be
\nabla \mu^{\pm}(\rr,t) -\FF^{\pm}(\rr) +e z^\pm \nabla  \psi(\rr,t)
\approx \FF^{\pm,drag} (\rr,t) .
\label{momentcomponent2}
\ee
Finally, using  eq. \eqref{dragforce}  we obtain
an  expression for the  ionic currents in terms of the microscopic parameters
which has the same form as the phenomenological Planck-Nernst current \eqref{microcurrent}, with the full chemical potential gradient 
$\mu^\pm$ replacing 
the ideal gas chemical potential gradient, $k_B T \nabla \ln \na $ 
 The total electric charge density current is
\be
{\bf J_e}=
 -\sum_\pm \frac{e z^\pm}{\gamma^\pm}  n^\pm(\rr,t)\nabla \mu^\pm(\rr,t) +  \sigma_{el} {\bf E}  +\rho_e(\rr,t) \uu(\rr,t)
\ee
where the zero frequency electric conductivity $\sigma_{el}$  for a uniform system is given by the Drude-Lorentz-like formula:
\be
\sigma_{el}=e^2  \Bigl( \frac{(z^+)^2}{\gamma^+ }+   \frac{(z^-)^2}{\gamma^- } \Bigl)
\label{Drude}
\ee
showing that the conductivity is due to collisions with the solvent and decreases as the solvent becomes denser
($\gamma^\pm$ is an increasing function of $n^0=n^+=n^-$) while increases with the number of charge carriers.

Finally we obtain 
the total momentum equation
\begin{align}
&\partial_{t}u_j(\rr,t)+ u_i(\rr,t)\nabla_i u_j(\rr,t)+\frac{1}{\rho_m}\nabla_i P^{(K)}_{ij}
+\frac{1}{\rho_m}\sum_\pm e z^\pm n^\pm(\rr,t)\nabla_j \psi(\rr,t)
\nonumber\\
& -\frac{1}{\rho_m} \sum_{\alpha=0,\pm} n^\alpha(\rr,t) \Bigl(F^{\alpha }_j(\rr) +F^{\alpha,mf}_{j}(\rr,t) +F^{\alpha,visc}_{j}(\rr,t)  \Bigl)=0.
\label{globalmomentumcont}
\end{align}
The total kinetic pressure which can be approximated as:
\be
\nabla P^{(K)}_{ij} \simeq k_B T \delta_{ij}\nabla_j \sum_\alpha n^\alpha(\rr,t)
 -\eta^{(K)}\Bigl(\frac{1}{3}\nabla_i\nabla_j u_i +\nabla_i^2 u_j\Bigl) ,
\ee
with $\eta^{(K)}=\frac{k_B T}{\nu}\sum_\alpha\na$.
Using the result of ref. \cite{marconi2009kinetic}, we can write
\be
\sum_\alpha\na(\rr,t)\FF^{\alpha,visc}(\rr,t)    
\simeq-\eta^{(C)}\nabla^2\uu-(\frac{1}{3}\eta^{(C)}+\eta_b^{(C)})\nabla(\nabla\cdot\uu).
\ee
The non-ideal contribution to the shear viscosity is  evaluated in uniform bulk conditions to be
\be
\eta^{(C)}=\frac{4}{15}\sum_{\alpha\beta}\sqrt{2\pi\muab k_B T}(\sab)^4 \gab \na_0\nb_0 ,
\ee

In conclusion,  \eqref{globalmomentumcont} can be cast in the Navier-Stokes form
\be
\partial_{t}u_j+ u_i\nabla_i u_j=-\frac{1}{\rho_m}\nabla_i P \delta_{ij}
-\frac{\rho_e}{\rho_m}\nabla_j \psi +\frac{\eta}{\rho_m} \nabla_i\nabla_i u_j
+\frac{\frac{4}{3}\eta}{\rho_m} \nabla_j \nabla_i u_i
\label{globalmomentumcont2}
\ee
with $\eta=\eta^{(K)}+\eta^{(C)}$ and $\nabla_i P=\nabla_i P_{id}+\sum_\alpha n^\alpha(\rr,t)\nabla_i \mu^{\alpha}_{exc}(\rr,t)$.
Eqs.  \eqref{microcurrent} together with the continuity equation for each species,  equation \eqref{globalmomentumcont2} 
and the Poisson equation are sufficient to understand the behavior of an electrolyte solution.

In refs.  \cite{melchionna2011electro,marini2012charge}
we have applied these equations to the study of electrokinetics flow in channels of nanometric section
and obtained the current-voltage relations and the conductance. In addition we have also considered the electro-osmosis phenomenon by which an electric field can be used to induce a mass flow in an electrolyte solution even in the absence of pressure gradients.

\section{Numerical  solution of the kinetic equation via LBM}
\label{numerical}

 
  In this section we briefly illustrate how the LBM is applied as a numerical solver to the
  present kinetic model, which  we shall discuss by considering only 
the one component case, while the multicomponent case can be easily deduced.
   Other popular schemes such as BGK  or Shan-Chen collisional forms 
  require a minor numerical effort, but on the other hand fail to predict 
  non trivial transport coefficients \cite{shan1994simulation}.

The WDLBM consists in integrating directly the kinetic equation equation \eqref{equilproc2} for $f(\rr,\vv,t)$ using a discretization procedure introduced
 some 20 years ago in fluid dynamics \cite{succi2001lattice,sukop2007lattice,guo2003discrete,guo2007discrete}. The LBM differs 
from other methods that are based on the solution of the
 set of equations for the density and the velocity of the fluid or from the Direct Monte Carlo simulation \cite{alexander1997direct}.
  The numerical method we have adopted is a substantial modification of the conventional method used in fluid dynamics 
applications to the presence of hard sphere collisions.
  
  The strategy behind the LBM is to
  reduce the number of possible particle spatial positions and microscopic velocities $\vv$
from a continuum to just a finite number of values,  ${\bf c}_{p}$, and
 similarly  discretize time into distinct steps.
The typical evolution equation is a time-explicit integration that reads
\begin{equation}
\frac{\partial f}{\partial t}+\vv\cdot \nabla f=\Omega'(f,M)
\label{bolt1}
\end{equation}
 where the kernel $\Omega'$ contains both the collisional term,
the BGK term and the external force term ${\bf F}\cdot\frac{\partial}{\partial {\vv}} f$,
while $M$ represents a generic moment of $f$.

We begin by projecting  the phase space distribution function 
over an orthonormal  basis spanned by the tensorial Hermite polynomials 
$\{H^{(l)}_{\underline\alpha}\}$:
\begin{equation}
f(\rr,\vv,t)=\omega(\vv)\sum_{l=0}^{\infty}\frac{1}{ l! v_{T}^{ 2l} } M^{(l)}_{\underline\alpha}(\rr,t)H^{(l)}_{\underline\alpha}(\vv)
\label{sumhermite}
\end{equation}
 where $\omega(\vv)=(2\pi v_{T}^{2})^{-3/2}e^{-\vv^{2}/2v_{T}^{2}}$
 and,
using the orthonormal relation 
 \be
  \int  \omega(\vv)  H^{(l)}_{\underline\alpha}(\vv)  H^{(m)}_{\underline\beta}(\vv)  d\vv=
  (v_T)^{l+m}\delta_{lm}\delta_{\underline\alpha \underline\beta}  
  \ee
the moments $M^{(l)}_{\underline\alpha}(\rr,t)$ can be obtained by:
 \be
 M^{(l)}_{\underline\alpha}(\rr,t)= \int  {f}(\rr,\vv,t) H^{(l)}_{\underline\alpha}(\vv)  d\vv
 \ee
 The exact infinite series  representation  given by eq. \eqref{sumhermite} is approximated
 by a function  $\bar{f}(\rr,\vv,t)$  obtained  by retaining in eq.(\ref{sumhermite}) only 
terms  up to  $l = K$.
Using the Gauss-Hermite quadrature formula 
in order to evaluate the expansion coefficients, $M^{(l)}_{\underline\alpha}(\rr,t)$, 
and using the nodes, ${\bf c}_{p}$,  and  the weights  $w_{p}$ of a quadrature of order 
$2G\geq K$
one obtains
\begin{eqnarray}
M^{(l)}_{\underline\alpha}(\rr,t) =\sum_{p=0}^{G}f_{p}(\rr,t)H^{(l)}_{\underline\alpha}({\bf c}_{p})
\end{eqnarray}
with
\be
f_{p}(\rr,t)\equiv \bar{f}(\rr,{\bf c}_{p},t)\frac{w_{p}}{\omega({\bf c}_{p})}
\ee
and $\bar {f}(\rr,\vv,t)\rightarrow f_p(\rr,t)$ is the truncation of order $K$ of the 
expansion of eq. \eqref{sumhermite}.  
  
  The distribution function is replaced by
an array of $Q$ populations ($19$ in the standard three dimensional case) and the distribution function  $f(\rr,\vv,t)\rightarrow f_p(\rr,t)$.
The propagation of the populations is achieved via a time discretization
to first order and a forward Euler update:
\be
\partial_{t}f_{p}(\rr,t)+\cc_{p}\cdot\partial_{\rr}f_{p}(\rr,t) \simeq
\frac{f_{p}(\rr+\cc_{p}\Delta t,t+\Delta t)-f_{p}(\rr,t)}{\Delta t}
\ee
where $\Delta t$ is the time-step.

 By the same token, we consider the expansion of
the collisional kernel, 
\begin{equation}
\Omega(\rr,\vv,t)=\omega(\vv)\sum_{l=0}^{\infty}\frac{1}{ l! v_{T}^{ 2l} } O^{(l)}_{\underline\alpha}(\rr,t)H^{(l)}_{\underline\alpha}(\vv)
\label{sumhermiteomega}
\end{equation}

The operational version of the collisional term is provided by 
\begin{equation}
\Omega_{p}(\rr,t)=\overline{\Omega}(\rr,{\bf c}_{p},t)\frac{w_{p}}{\omega(c_{p})}
\end{equation}
From these transformations, the evolution equation of the new representation
is given by the following updating scheme 
\begin{equation}
{f}_{p}(\rr+\cc_{p}\Delta t,t+\Delta t)={f}_{p}(\rr,t)+\Omega_{p}(\rr,t)\Delta t
\label{eq:newevolution}
\end{equation}
where
\begin{eqnarray}
\Omega_{p}(\rr,t) & = & \nu(f_{p}^{eq}(\rr,t)-f_{p}(\rr,t))+S_{p}(\rr,t)
\label{eq:kernelf}
\end{eqnarray}

The explicit form of the r.h.s. of eqs. \eqref{eq:kernelf}-\eqref{eq:kernelg} reads 
\begin{eqnarray}
f_{p}^{eq} (\rr,t)& = & w_{p}\left[n(\rr,t)+n(\rr,t) \uu(\rr,t)\cdot{\cal H}_{p}^{(1)}(\cc_p)+n(\rr,t)
\uu(\rr,t)\uu(\rr,t):{\cal H}_{p}^{(2)}(\cc_p)\right]
\label{eq:kernelf}\\
S_{p}(\rr,t) & = & 
w_{p} \left\{ 
  n(\rr,t)\left({\bf F}^{mf}(\rr,t)+{\bf F}^{visc}(\rr,t)+\FF^{ext}(\rr)\right)\cdot
\left[{\cal H}_{p}^{(1)}(\cc_p)+2{\cal H}_{p}^{(2)}(\cc_p)\cdot\uu(\rr,t)\right]\right\}
\label{eq:kernelg}
\end{eqnarray}
 where ${\cal H}^{(1)}(\cc_{p})=\frac{\cc_{p}}{v_{T}^{2}}$
and ${\cal H}^{(2)}(\cc_{p})=\frac{\cc_{p}\cc_{p}-v_{T}^{2}{\bf I}}{2v_{T}^{4}}$,
being a vector and a tensor of rank two, respectively, and ${\bf I}$ is the unit tensor.

Once the populations $f_p$ are known, they are used to compute hydrodynamic moments, 
entering both the equilibrium and in sampling the macroscopic evolution.
The fluid density, momentum current and local temperature read
\bea
n(\rr,t) &=& \sum_p f_p(\rr,t) \nonumber \\
n(\rr,t) \uu(\rr,t) &=& \sum_p f_p(\rr,t) \cc_p \nonumber \\
\frac{3 k_B}{m} n(\rr,t) T(\rr,t) &=& \sum_p f_p(\rr,t) \cc_p^2
\eea

The dynamics of the charged multicomponent system can be solved numerically in the same framework
by generalizing the procedure described above to three species.
The novelty consists in the presence of the electric field $\psi(\rr,t)$ which must be
determined from the charge distribution by solving the   
Poisson equation for the electrostatic potential generated by the  mobile and surface charges. 
The determination of $\psi(\rr)$ can be achieved by using a successive over-relaxation method,
while the speed of convergence is greatly enhanced by employing a 
Gauss-Seidel checker-board scheme in conjunction with Chebychev acceleration \cite{press1992numerical}. 
Neumann boundary conditions on the gradient of the electrostatic potential are imposed at the wall surface
\be
\hat n \cdot \nabla \psi |_{\rr \in S}=-\frac{\Sigma(\rr)}{\epsilon}
\ee
where $\hat n$ is the normal to the surface, $S$.

\section{Conclusions}
\label{conclusions}
In summary, we reviewed recent advances in kinetic theory and modeling applied to the
transport of molecular liquids confined in very small spaces.
We formulated the model and then considered the series of
approximations needed in order to achieve a workable numerical scheme to be used under 
generic confinement conditions.

At first, we  split the collision operator into  hydrodynamic and  non-hydrodynamic 
(i.e. purely kinetic) contributions. The non-hydrodynamic contribution
is handled by the BGK ansatz, which gives rise to the ideal gas
thermodynamics and to non-collisional stress terms.
The non hydrodynamic part instead is treated by performing the integrals featuring in the collision operator
by using a parametric form of the phase space distribution $f$. 
Since the collision operator involves convolutions with the configurational pair correlation function
we used the so-called adiabatic approximation to determine it.
The resulting equation involves only the single particle distribution function and its first few moments,
and can be solved by via the Lattice Boltzmann method. Such strategy gives access to the complete hydrodynamic 
information without the need to solve the hydrodynamic equations separately.  
\bibliographystyle{apsrev}
\bibliography{marconibeijing2013} 			     

\providecommand{\noopsort}[1]{}\providecommand{\singleletter}[1]{#1}%
\begin{thebibliography}{52}
\expandafter\ifx\csname natexlab\endcsname\relax\def\natexlab#1{#1}\fi
\expandafter\ifx\csname bibnamefont\endcsname\relax
  \def\bibnamefont#1{#1}\fi
\expandafter\ifx\csname bibfnamefont\endcsname\relax
  \def\bibfnamefont#1{#1}\fi
\expandafter\ifx\csname citenamefont\endcsname\relax
  \def\citenamefont#1{#1}\fi
\expandafter\ifx\csname url\endcsname\relax
  \def\url#1{\texttt{#1}}\fi
\expandafter\ifx\csname urlprefix\endcsname\relax\def\urlprefix{URL }\fi
\providecommand{\bibinfo}[2]{#2}
\providecommand{\eprint}[2][]{\url{#2}}

\bibitem[{\citenamefont{Castiglione et~al.}(2008)\citenamefont{Castiglione,
  Falcioni, Lesne, and Vulpiani}}]{castiglione2008chaos}
\bibinfo{author}{\bibfnamefont{P.}~\bibnamefont{Castiglione}},
  \bibinfo{author}{\bibfnamefont{M.}~\bibnamefont{Falcioni}},
  \bibinfo{author}{\bibfnamefont{A.}~\bibnamefont{Lesne}}, \bibnamefont{and}
  \bibinfo{author}{\bibfnamefont{A.}~\bibnamefont{Vulpiani}},
  \emph{\bibinfo{title}{Chaos and coarse graining in statistical mechanics}}
  (\bibinfo{publisher}{Cambridge University Press Cambridge},
  \bibinfo{year}{2008}).

\bibitem[{\citenamefont{Pavliotis and Stuart}(2008)}]{pavliotis2008multiscale}
\bibinfo{author}{\bibfnamefont{G.}~\bibnamefont{Pavliotis}} \bibnamefont{and}
  \bibinfo{author}{\bibfnamefont{A.}~\bibnamefont{Stuart}},
  \emph{\bibinfo{title}{Multiscale methods: averaging and homogenization}},
  vol.~\bibinfo{volume}{53} (\bibinfo{publisher}{Springer},
  \bibinfo{year}{2008}).

\bibitem[{\citenamefont{Weinan}(2011)}]{weinan2011principles}
\bibinfo{author}{\bibfnamefont{E.}~\bibnamefont{Weinan}},
  \emph{\bibinfo{title}{Principles of multiscale modeling}}
  (\bibinfo{publisher}{Cambridge University Press}, \bibinfo{year}{2011}).

\bibitem[{\citenamefont{Huang}(2001)}]{huang2001introduction}
\bibinfo{author}{\bibfnamefont{K.}~\bibnamefont{Huang}},
  \emph{\bibinfo{title}{Introduction to statistical physics}}
  (\bibinfo{publisher}{CRC Press}, \bibinfo{year}{2001}).

\bibitem[{\citenamefont{Batchelor}(2000)}]{batchelor2000introduction}
\bibinfo{author}{\bibfnamefont{G.~K.} \bibnamefont{Batchelor}},
  \emph{\bibinfo{title}{An introduction to fluid dynamics}}
  (\bibinfo{publisher}{Cambridge university press}, \bibinfo{year}{2000}).

\bibitem[{\citenamefont{Marconi and Tarazona}(1999)}]{marconi1999dynamic}
\bibinfo{author}{\bibfnamefont{U.~M.~B.} \bibnamefont{Marconi}}
  \bibnamefont{and} \bibinfo{author}{\bibfnamefont{P.}~\bibnamefont{Tarazona}},
  \bibinfo{journal}{The Journal of Chemical Physics}
  \textbf{\bibinfo{volume}{110}}, \bibinfo{pages}{8032} (\bibinfo{year}{1999}).

\bibitem[{\citenamefont{Archer and Evans}(2004)}]{archer2004dynamical}
\bibinfo{author}{\bibfnamefont{A.}~\bibnamefont{Archer}} \bibnamefont{and}
  \bibinfo{author}{\bibfnamefont{R.}~\bibnamefont{Evans}},
  \bibinfo{journal}{The Journal of Chemical Physics}
  \textbf{\bibinfo{volume}{121}}, \bibinfo{pages}{4246} (\bibinfo{year}{2004}).

\bibitem[{\citenamefont{Rauscher}(2010)}]{rauscher2010ddft}
\bibinfo{author}{\bibfnamefont{M.}~\bibnamefont{Rauscher}},
  \bibinfo{journal}{Journal of Physics: Condensed Matter}
  \textbf{\bibinfo{volume}{22}}, \bibinfo{pages}{364109}
  (\bibinfo{year}{2010}).

\bibitem[{\citenamefont{Marini Bettolo~Marconi and
  Melchionna}(2009)}]{marconi2009kinetic}
\bibinfo{author}{\bibfnamefont{U.}~\bibnamefont{Marini Bettolo~Marconi}}
  \bibnamefont{and}
  \bibinfo{author}{\bibfnamefont{S.}~\bibnamefont{Melchionna}},
  \bibinfo{journal}{The Journal of Chemical Physics}
  \textbf{\bibinfo{volume}{131}}, \bibinfo{pages}{014105}
  (\bibinfo{year}{2009}).

\bibitem[{\citenamefont{Marconi and Melchionna}(2010)}]{marconi2010dynamic}
\bibinfo{author}{\bibfnamefont{U.~M.~B.} \bibnamefont{Marconi}}
  \bibnamefont{and}
  \bibinfo{author}{\bibfnamefont{S.}~\bibnamefont{Melchionna}},
  \bibinfo{journal}{Journal of Physics: Condensed Matter}
  \textbf{\bibinfo{volume}{22}}, \bibinfo{pages}{364110}
  (\bibinfo{year}{2010}).

\bibitem[{\citenamefont{Shan and Chen}(1994)}]{shan1994simulation}
\bibinfo{author}{\bibfnamefont{X.}~\bibnamefont{Shan}} \bibnamefont{and}
  \bibinfo{author}{\bibfnamefont{H.}~\bibnamefont{Chen}},
  \bibinfo{journal}{Physical Review E} \textbf{\bibinfo{volume}{49}},
  \bibinfo{pages}{2941} (\bibinfo{year}{1994}).

\bibitem[{\citenamefont{He et~al.}(1998)\citenamefont{He, Shan, and
  Doolen}}]{he1998discrete}
\bibinfo{author}{\bibfnamefont{X.}~\bibnamefont{He}},
  \bibinfo{author}{\bibfnamefont{X.}~\bibnamefont{Shan}}, \bibnamefont{and}
  \bibinfo{author}{\bibfnamefont{G.~D.} \bibnamefont{Doolen}},
  \bibinfo{journal}{Physical Review E} \textbf{\bibinfo{volume}{57}},
  \bibinfo{pages}{R13} (\bibinfo{year}{1998}).

\bibitem[{\citenamefont{Risken}(1984)}]{risken1984fokker}
\bibinfo{author}{\bibfnamefont{H.}~\bibnamefont{Risken}},
  \emph{\bibinfo{title}{Fokker-Planck Equation}}
  (\bibinfo{publisher}{Springer}, \bibinfo{year}{1984}).

\bibitem[{\citenamefont{Hansen and McDonald}(1990)}]{hansen1990theory}
\bibinfo{author}{\bibfnamefont{J.-P.} \bibnamefont{Hansen}} \bibnamefont{and}
  \bibinfo{author}{\bibfnamefont{I.~R.} \bibnamefont{McDonald}},
  \emph{\bibinfo{title}{Theory of simple liquids}}
  (\bibinfo{publisher}{Elsevier}, \bibinfo{year}{1990}).

\bibitem[{\citenamefont{Dufty and Baskaran}(2005)}]{dufty2005hard}
\bibinfo{author}{\bibfnamefont{J.~W.} \bibnamefont{Dufty}} \bibnamefont{and}
  \bibinfo{author}{\bibfnamefont{A.}~\bibnamefont{Baskaran}},
  \bibinfo{journal}{Annals of the New York Academy of Sciences}
  \textbf{\bibinfo{volume}{1045}}, \bibinfo{pages}{93} (\bibinfo{year}{2005}).

\bibitem[{\citenamefont{Brilliantov and
  P{\"o}schel}(2010)}]{brilliantov2010kinetic}
\bibinfo{author}{\bibfnamefont{N.~V.} \bibnamefont{Brilliantov}}
  \bibnamefont{and}
  \bibinfo{author}{\bibfnamefont{T.}~\bibnamefont{P{\"o}schel}},
  \emph{\bibinfo{title}{Kinetic theory of granular gases}}
  (\bibinfo{publisher}{Oxford University Press}, \bibinfo{year}{2010}).

\bibitem[{\citenamefont{Van~Beijeren and
  Ernst}(1973{\natexlab{a}})}]{van1973modified}
\bibinfo{author}{\bibfnamefont{H.}~\bibnamefont{Van~Beijeren}}
  \bibnamefont{and} \bibinfo{author}{\bibfnamefont{M.~H.} \bibnamefont{Ernst}},
  \bibinfo{journal}{Physica} \textbf{\bibinfo{volume}{68}},
  \bibinfo{pages}{437} (\bibinfo{year}{1973}{\natexlab{a}}).

\bibitem[{\citenamefont{Van~Beijeren and
  Ernst}(1973{\natexlab{b}})}]{van1973modified2}
\bibinfo{author}{\bibfnamefont{H.}~\bibnamefont{Van~Beijeren}}
  \bibnamefont{and} \bibinfo{author}{\bibfnamefont{M.}~\bibnamefont{Ernst}},
  \bibinfo{journal}{Physica} \textbf{\bibinfo{volume}{70}},
  \bibinfo{pages}{225} (\bibinfo{year}{1973}{\natexlab{b}}).

\bibitem[{\citenamefont{Marini Bettolo~Marconi and
  Melchionna}(2007)}]{marconi2007phase}
\bibinfo{author}{\bibfnamefont{U.}~\bibnamefont{Marini Bettolo~Marconi}}
  \bibnamefont{and}
  \bibinfo{author}{\bibfnamefont{S.}~\bibnamefont{Melchionna}},
  \bibinfo{journal}{The Journal of Chemical Physics}
  \textbf{\bibinfo{volume}{126}}, \bibinfo{pages}{184109}
  (\bibinfo{year}{2007}).

\bibitem[{\citenamefont{Fischer and Methfessel}(1980)}]{fischer1980born}
\bibinfo{author}{\bibfnamefont{J.}~\bibnamefont{Fischer}} \bibnamefont{and}
  \bibinfo{author}{\bibfnamefont{M.}~\bibnamefont{Methfessel}},
  \bibinfo{journal}{Physical Review A} \textbf{\bibinfo{volume}{22}},
  \bibinfo{pages}{2836} (\bibinfo{year}{1980}).

\bibitem[{\citenamefont{Evans}(1979)}]{evans1979nature}
\bibinfo{author}{\bibfnamefont{R.}~\bibnamefont{Evans}},
  \bibinfo{journal}{Advances in Physics} \textbf{\bibinfo{volume}{28}},
  \bibinfo{pages}{143} (\bibinfo{year}{1979}).

\bibitem[{\citenamefont{Marconi and
  Tarazona}(2006)}]{marconi2006nonequilibrium}
\bibinfo{author}{\bibfnamefont{U.~M.~B.} \bibnamefont{Marconi}}
  \bibnamefont{and} \bibinfo{author}{\bibfnamefont{P.}~\bibnamefont{Tarazona}},
  \bibinfo{journal}{The Journal of Chemical Physics}
  \textbf{\bibinfo{volume}{124}}, \bibinfo{pages}{164901}
  (\bibinfo{year}{2006}).

\bibitem[{\citenamefont{Marini-Bettolo-Marconi
  et~al.}(2007)\citenamefont{Marini-Bettolo-Marconi, Tarazona, and
  Cecconi}}]{marini2007theory}
\bibinfo{author}{\bibfnamefont{U.}~\bibnamefont{Marini-Bettolo-Marconi}},
  \bibinfo{author}{\bibfnamefont{P.}~\bibnamefont{Tarazona}}, \bibnamefont{and}
  \bibinfo{author}{\bibfnamefont{F.}~\bibnamefont{Cecconi}},
  \bibinfo{journal}{The Journal of chemical physics}
  \textbf{\bibinfo{volume}{126}}, \bibinfo{pages}{164904}
  (\bibinfo{year}{2007}).

\bibitem[{\citenamefont{Goddard et~al.}(2012)\citenamefont{Goddard, Nold,
  Savva, Pavliotis, and Kalliadasis}}]{goddard2012general}
\bibinfo{author}{\bibfnamefont{B.~D.} \bibnamefont{Goddard}},
  \bibinfo{author}{\bibfnamefont{A.}~\bibnamefont{Nold}},
  \bibinfo{author}{\bibfnamefont{N.}~\bibnamefont{Savva}},
  \bibinfo{author}{\bibfnamefont{G.~A.} \bibnamefont{Pavliotis}},
  \bibnamefont{and}
  \bibinfo{author}{\bibfnamefont{S.}~\bibnamefont{Kalliadasis}},
  \bibinfo{journal}{Physical Review Letters} \textbf{\bibinfo{volume}{109}},
  \bibinfo{pages}{120603} (\bibinfo{year}{2012}).

\bibitem[{\citenamefont{Wilemski}(1976)}]{wilemski1976derivation}
\bibinfo{author}{\bibfnamefont{G.}~\bibnamefont{Wilemski}},
  \bibinfo{journal}{Journal of Statistical Physics}
  \textbf{\bibinfo{volume}{14}}, \bibinfo{pages}{153} (\bibinfo{year}{1976}).

\bibitem[{\citenamefont{Titulaer}(1978)}]{titulaer1978systematic}
\bibinfo{author}{\bibfnamefont{U.~M.} \bibnamefont{Titulaer}},
  \bibinfo{journal}{Physica A: Statistical Mechanics and its Applications}
  \textbf{\bibinfo{volume}{91}}, \bibinfo{pages}{321} (\bibinfo{year}{1978}).

\bibitem[{\citenamefont{Anero and Espa{\~n}ol}(2007)}]{anero2007dynamic}
\bibinfo{author}{\bibfnamefont{J.}~\bibnamefont{Anero}} \bibnamefont{and}
  \bibinfo{author}{\bibfnamefont{P.}~\bibnamefont{Espa{\~n}ol}},
  \bibinfo{journal}{EPL (Europhysics Letters)} \textbf{\bibinfo{volume}{78}},
  \bibinfo{pages}{50005} (\bibinfo{year}{2007}).

\bibitem[{\citenamefont{Anero et~al.}(2013)\citenamefont{Anero, Espa{\~n}ol,
  and Tarazona}}]{anero2013functional}
\bibinfo{author}{\bibfnamefont{J.~G.} \bibnamefont{Anero}},
  \bibinfo{author}{\bibfnamefont{P.}~\bibnamefont{Espa{\~n}ol}},
  \bibnamefont{and} \bibinfo{author}{\bibfnamefont{P.}~\bibnamefont{Tarazona}},
  \bibinfo{journal}{The Journal of chemical physics}
  \textbf{\bibinfo{volume}{139}}, \bibinfo{pages}{034106}
  (\bibinfo{year}{2013}).

\bibitem[{\citenamefont{Archer}(2009)}]{archer2009dynamical}
\bibinfo{author}{\bibfnamefont{A.}~\bibnamefont{Archer}}, \bibinfo{journal}{The
  Journal of chemical physics} \textbf{\bibinfo{volume}{130}},
  \bibinfo{pages}{014509} (\bibinfo{year}{2009}).

\bibitem[{\citenamefont{Dufty et~al.}(1996)\citenamefont{Dufty, Santos, and
  Brey}}]{dufty1996practical}
\bibinfo{author}{\bibfnamefont{J.~W.} \bibnamefont{Dufty}},
  \bibinfo{author}{\bibfnamefont{A.}~\bibnamefont{Santos}}, \bibnamefont{and}
  \bibinfo{author}{\bibfnamefont{J.~J.} \bibnamefont{Brey}},
  \bibinfo{journal}{Physical review letters} \textbf{\bibinfo{volume}{77}},
  \bibinfo{pages}{1270} (\bibinfo{year}{1996}).

\bibitem[{\citenamefont{Santos et~al.}(1998)\citenamefont{Santos, Montanero,
  Dufty, and Brey}}]{santos1998kinetic}
\bibinfo{author}{\bibfnamefont{A.}~\bibnamefont{Santos}},
  \bibinfo{author}{\bibfnamefont{J.~M.} \bibnamefont{Montanero}},
  \bibinfo{author}{\bibfnamefont{J.~W.} \bibnamefont{Dufty}}, \bibnamefont{and}
  \bibinfo{author}{\bibfnamefont{J.~J.} \bibnamefont{Brey}},
  \bibinfo{journal}{Physical Review E} \textbf{\bibinfo{volume}{57}},
  \bibinfo{pages}{1644} (\bibinfo{year}{1998}).

\bibitem[{\citenamefont{Bhatnagar et~al.}(1954)\citenamefont{Bhatnagar, Gross,
  and Krook}}]{bhatnagar1954model}
\bibinfo{author}{\bibfnamefont{P.~L.} \bibnamefont{Bhatnagar}},
  \bibinfo{author}{\bibfnamefont{E.~P.} \bibnamefont{Gross}}, \bibnamefont{and}
  \bibinfo{author}{\bibfnamefont{M.}~\bibnamefont{Krook}},
  \bibinfo{journal}{Physical review} \textbf{\bibinfo{volume}{94}},
  \bibinfo{pages}{511} (\bibinfo{year}{1954}).

\bibitem[{\citenamefont{Longuet-Higgins and
  Pople}(1956)}]{longuet1956transport}
\bibinfo{author}{\bibfnamefont{H.}~\bibnamefont{Longuet-Higgins}}
  \bibnamefont{and} \bibinfo{author}{\bibfnamefont{J.}~\bibnamefont{Pople}},
  \bibinfo{journal}{The Journal of Chemical Physics}
  \textbf{\bibinfo{volume}{25}}, \bibinfo{pages}{884} (\bibinfo{year}{1956}).

\bibitem[{\citenamefont{Marconi and Melchionna}(2013)}]{marconi2013weighted}
\bibinfo{author}{\bibfnamefont{U.~M.~B.} \bibnamefont{Marconi}}
  \bibnamefont{and}
  \bibinfo{author}{\bibfnamefont{S.}~\bibnamefont{Melchionna}},
  \bibinfo{journal}{Molecular Physics} \textbf{\bibinfo{volume}{111}},
  \bibinfo{pages}{3126} (\bibinfo{year}{2013}).

\bibitem[{\citenamefont{Luo et~al.}(2011)\citenamefont{Luo, Liao, Chen, Peng,
  and Zhang}}]{luo2011numerics}
\bibinfo{author}{\bibfnamefont{L.-S.} \bibnamefont{Luo}},
  \bibinfo{author}{\bibfnamefont{W.}~\bibnamefont{Liao}},
  \bibinfo{author}{\bibfnamefont{X.}~\bibnamefont{Chen}},
  \bibinfo{author}{\bibfnamefont{Y.}~\bibnamefont{Peng}}, \bibnamefont{and}
  \bibinfo{author}{\bibfnamefont{W.}~\bibnamefont{Zhang}},
  \bibinfo{journal}{Physical Review E} \textbf{\bibinfo{volume}{83}},
  \bibinfo{pages}{056710} (\bibinfo{year}{2011}).

\bibitem[{\citenamefont{Luo et~al.}(2010)\citenamefont{Luo, Krafczyk, and
  Shyy}}]{luo2010lattice}
\bibinfo{author}{\bibfnamefont{L.-S.} \bibnamefont{Luo}},
  \bibinfo{author}{\bibfnamefont{M.}~\bibnamefont{Krafczyk}}, \bibnamefont{and}
  \bibinfo{author}{\bibfnamefont{W.}~\bibnamefont{Shyy}},
  \bibinfo{journal}{Encyclopedia of Aerospace Engineering}
  (\bibinfo{year}{2010}).

\bibitem[{\citenamefont{Baskaran et~al.}(2013)\citenamefont{Baskaran, Baskaran,
  and Lowengrub}}]{baskaran2013kinetic}
\bibinfo{author}{\bibfnamefont{A.}~\bibnamefont{Baskaran}},
  \bibinfo{author}{\bibfnamefont{A.}~\bibnamefont{Baskaran}}, \bibnamefont{and}
  \bibinfo{author}{\bibfnamefont{J.}~\bibnamefont{Lowengrub}},
  \bibinfo{journal}{arXiv preprint arXiv:1310.6070}  (\bibinfo{year}{2013}).

\bibitem[{\citenamefont{Oleksy and Hansen}(2009)}]{oleksy2009microscopic}
\bibinfo{author}{\bibfnamefont{A.}~\bibnamefont{Oleksy}} \bibnamefont{and}
  \bibinfo{author}{\bibfnamefont{J.-P.} \bibnamefont{Hansen}},
  \bibinfo{journal}{Molecular Physics} \textbf{\bibinfo{volume}{107}},
  \bibinfo{pages}{2609} (\bibinfo{year}{2009}).

\bibitem[{\citenamefont{Marini Bettolo~Marconi and
  Melchionna}(2012)}]{marini2012charge}
\bibinfo{author}{\bibfnamefont{U.}~\bibnamefont{Marini Bettolo~Marconi}}
  \bibnamefont{and}
  \bibinfo{author}{\bibfnamefont{S.}~\bibnamefont{Melchionna}},
  \bibinfo{journal}{Langmuir} \textbf{\bibinfo{volume}{28}},
  \bibinfo{pages}{13727} (\bibinfo{year}{2012}).

\bibitem[{\citenamefont{Marini Bettolo~Marconi and
  Melchionna}(2011{\natexlab{a}})}]{marconi2011multicomponent}
\bibinfo{author}{\bibfnamefont{U.}~\bibnamefont{Marini Bettolo~Marconi}}
  \bibnamefont{and}
  \bibinfo{author}{\bibfnamefont{S.}~\bibnamefont{Melchionna}},
  \bibinfo{journal}{The Journal of Chemical Physics}
  \textbf{\bibinfo{volume}{135}}, \bibinfo{pages}{044104}
  (\bibinfo{year}{2011}{\natexlab{a}}).

\bibitem[{\citenamefont{Marconi}(2011)}]{marconi2011non}
\bibinfo{author}{\bibfnamefont{U.~M.~B.} \bibnamefont{Marconi}},
  \bibinfo{journal}{Molecular Physics} \textbf{\bibinfo{volume}{109}},
  \bibinfo{pages}{1265} (\bibinfo{year}{2011}).

\bibitem[{\citenamefont{Boubl{\'\i}k}(1970)}]{boublik1970hard}
\bibinfo{author}{\bibfnamefont{T.}~\bibnamefont{Boubl{\'\i}k}},
  \bibinfo{journal}{The Journal of Chemical Physics}
  \textbf{\bibinfo{volume}{53}}, \bibinfo{pages}{471} (\bibinfo{year}{1970}).

\bibitem[{\citenamefont{Mansoori et~al.}(1971)\citenamefont{Mansoori, Carnahan,
  Starling, and Leland~Jr}}]{mansoori1971equilibrium}
\bibinfo{author}{\bibfnamefont{G.}~\bibnamefont{Mansoori}},
  \bibinfo{author}{\bibfnamefont{N.}~\bibnamefont{Carnahan}},
  \bibinfo{author}{\bibfnamefont{K.}~\bibnamefont{Starling}}, \bibnamefont{and}
  \bibinfo{author}{\bibfnamefont{T.}~\bibnamefont{Leland~Jr}},
  \bibinfo{journal}{The Journal of Chemical Physics}
  \textbf{\bibinfo{volume}{54}}, \bibinfo{pages}{1523} (\bibinfo{year}{1971}).

\bibitem[{\citenamefont{Wendland}(1997)}]{wendland1997born}
\bibinfo{author}{\bibfnamefont{M.}~\bibnamefont{Wendland}},
  \bibinfo{journal}{Fluid phase equilibria} \textbf{\bibinfo{volume}{141}},
  \bibinfo{pages}{25} (\bibinfo{year}{1997}).

\bibitem[{\citenamefont{Marini Bettolo~Marconi and
  Melchionna}(2011{\natexlab{b}})}]{marconi2011dynamics}
\bibinfo{author}{\bibfnamefont{U.}~\bibnamefont{Marini Bettolo~Marconi}}
  \bibnamefont{and}
  \bibinfo{author}{\bibfnamefont{S.}~\bibnamefont{Melchionna}},
  \bibinfo{journal}{The Journal of Chemical Physics}
  \textbf{\bibinfo{volume}{134}}, \bibinfo{pages}{064118}
  (\bibinfo{year}{2011}{\natexlab{b}}).

\bibitem[{\citenamefont{Melchionna and Marini
  Bettolo~Marconi}(2011)}]{melchionna2011electro}
\bibinfo{author}{\bibfnamefont{S.}~\bibnamefont{Melchionna}} \bibnamefont{and}
  \bibinfo{author}{\bibfnamefont{U.}~\bibnamefont{Marini Bettolo~Marconi}},
  \bibinfo{journal}{EPL (EuroPhysics Letters)} \textbf{\bibinfo{volume}{95}},
  \bibinfo{pages}{44002} (\bibinfo{year}{2011}).

\bibitem[{\citenamefont{Succi}(2001)}]{succi2001lattice}
\bibinfo{author}{\bibfnamefont{S.}~\bibnamefont{Succi}},
  \emph{\bibinfo{title}{The lattice Boltzmann equation: for fluid dynamics and
  beyond}} (\bibinfo{publisher}{Oxford university press},
  \bibinfo{year}{2001}).

\bibitem[{\citenamefont{Sukop and Thorne}(2007)}]{sukop2007lattice}
\bibinfo{author}{\bibfnamefont{M.~C.} \bibnamefont{Sukop}} \bibnamefont{and}
  \bibinfo{author}{\bibfnamefont{D.~T.} \bibnamefont{Thorne}},
  \emph{\bibinfo{title}{Lattice Boltzmann modeling: an introduction for
  geoscientists and engineers}} (\bibinfo{publisher}{Springer},
  \bibinfo{year}{2007}).

\bibitem[{\citenamefont{Guo and Zhao}(2003)}]{guo2003discrete}
\bibinfo{author}{\bibfnamefont{Z.}~\bibnamefont{Guo}} \bibnamefont{and}
  \bibinfo{author}{\bibfnamefont{T.}~\bibnamefont{Zhao}},
  \bibinfo{journal}{Physical Review E} \textbf{\bibinfo{volume}{68}},
  \bibinfo{pages}{035302} (\bibinfo{year}{2003}).

\bibitem[{\citenamefont{Guo et~al.}(2007)\citenamefont{Guo, Shi, Zhao, and
  Zheng}}]{guo2007discrete}
\bibinfo{author}{\bibfnamefont{Z.}~\bibnamefont{Guo}},
  \bibinfo{author}{\bibfnamefont{B.}~\bibnamefont{Shi}},
  \bibinfo{author}{\bibfnamefont{T.-S.} \bibnamefont{Zhao}}, \bibnamefont{and}
  \bibinfo{author}{\bibfnamefont{C.}~\bibnamefont{Zheng}},
  \bibinfo{journal}{Physical Review E} \textbf{\bibinfo{volume}{76}},
  \bibinfo{pages}{056704} (\bibinfo{year}{2007}).

\bibitem[{\citenamefont{Alexander and Garcia}(1997)}]{alexander1997direct}
\bibinfo{author}{\bibfnamefont{F.~J.} \bibnamefont{Alexander}}
  \bibnamefont{and} \bibinfo{author}{\bibfnamefont{A.~L.}
  \bibnamefont{Garcia}}, \bibinfo{journal}{Computers in Physics}
  \textbf{\bibinfo{volume}{11}}, \bibinfo{pages}{588} (\bibinfo{year}{1997}).

\bibitem[{\citenamefont{Press et~al.}(1992)\citenamefont{Press, Flannery,
  Teukolsky, and Vetterling}}]{press1992numerical}
\bibinfo{author}{\bibfnamefont{W.~H.} \bibnamefont{Press}},
  \bibinfo{author}{\bibfnamefont{B.~P.} \bibnamefont{Flannery}},
  \bibinfo{author}{\bibfnamefont{S.~A.} \bibnamefont{Teukolsky}},
  \bibnamefont{and} \bibinfo{author}{\bibfnamefont{W.~T.}
  \bibnamefont{Vetterling}}, \emph{\bibinfo{title}{Numerical Recipes in FORTRAN
  77: Volume 1, Volume 1 of Fortran Numerical Recipes: The Art of Scientific
  Computing}}, vol.~\bibinfo{volume}{1} (\bibinfo{publisher}{Cambridge
  university press}, \bibinfo{year}{1992}).

\end{thebibliography}
\end{document}